\documentclass[prd,showpacs,showkeys,nofootinbib,floatfix,eqsecnum,
               fleqn,preprint,12pt,tightenlines]{revtex4-1} 

\usepackage{amsmath,amssymb,revsymb,graphicx,dcolumn}

\newcommand{\version}{v2} 

\newcommand{\beq}{\begin{equation}}
\newcommand{\eeq}{\end{equation}}
\newcommand{\beqa}{\begin{eqnarray}}
\newcommand{\eeqa}{\end{eqnarray}}
\newcommand{\bsubeqs}{\begin{subequations}}
\newcommand{\esubeqs}{\end{subequations}}

\begin{document}

\begin{widetext}
%
\noindent arXiv:1911.11116
\hfill
KA--TP--22--2019\;(\version)
%
\vspace*{3mm}
\end{widetext}

\title{Instability of the big bang coordinate singularity\\ in a
Milne-like universe}%

\vspace*{4mm}

\author{F.R. Klinkhamer}
\email{frans.klinkhamer@kit.edu}
\affiliation{Institute for Theoretical Physics,
Karlsruhe Institute of Technology (KIT),\\
76128 Karlsruhe, Germany}

\author{Z.L. Wang}
 \email{ziliang.wang@kit.edu}

 \affiliation{Institute for Theoretical Physics,
 Karlsruhe Institute of Technology (KIT),\\
 76128 Karlsruhe, Germany
 \vspace*{4mm}}

\begin{abstract}
\vspace*{1mm}\noindent
We present a simplified dynamic-vacuum-energy model
for a time-symmetric Milne-like universe.
The big bang singularity in this simplified model, like the one in a previous model, is just a coordinate singularity
with finite curvature and energy density.
We then calculate the dynamic behavior of scalar metric
perturbations and find that these perturbations
destabilize the big bang singularity.
\\
\vspace*{-0mm}
\end{abstract}


\maketitle

\section{Introduction}
\label{sec:Intro}

We have recently discussed two ways to ``tame'' the
big bang singularity of the Friedmann
solution~\cite{Friedmann1922-1924,Weinberg1972,MisnerThorneWheeler2017}.
The first way~\cite{Klinkhamer2019,KlinkhamerWang2019-PRD,%
KlinkhamerWang2019-LHEP,Klinkhamer2019-revisited,%
KlinkhamerWang2019-nonsing-bounce-pert}
is to consider a particular degenerate
metric, which describes  a spacetime defect with
a characteristic length scale $b$
(the characteristic time scale is then $b/c$,
where $c$ is the velocity of light in vacuum).
The cosmic scale factor $a(t)$ remains finite at
$t=t_\text{bb} \equiv 0$
and the singular big bang is turned into a nonsingular bounce.

The second way~\cite{Ling2018,KlinkhamerLing2019}
keeps a smooth spacetime
(with the spatially-hyperbolic Robertson--Walker metric)
but allows only vacuum energy to be present. There is still
the big bang at $t=t_\text{bb} \equiv 0$ with vanishing cosmic scale factor,
$a(0)=0$,
but there is now only a coordinate singularity with finite values
of the curvature scalars.
The behavior at $t=0$ resembles
that of the so-called Milne universe~\cite{Milne1932}
(cf. Sec.~5.3 of Ref.~\cite{BirrellDavies1982}
and Sec.~1.3.5 of Ref.~\cite{Mukhanov2005}),
which corresponds to Minkowski spacetime in expanding/contracting
coordinates. More precisely, the spacetime of
Refs.~\cite{Ling2018,KlinkhamerLing2019}
is called a ``Milne-like universe,'' as it corresponds,
at $t=0$, to a slice of de Sitter spacetime rather
than Minkowski spacetime.

The physics model of Ref.~\cite{KlinkhamerLing2019} relies
on the $q$-theory approach to the cosmological constant
problem~\cite{KlinkhamerVolovik2008a,KlinkhamerVolovik2008b,%
KlinkhamerVolovik2009,KlinkhamerVolovik2016} and shows that
the positive vacuum energy at $t=0$ reduces to zero as the
model universe evolves, $|t|\to\infty$.
But the explicit $q$-theory for the Milne-like
model of Ref.~\cite{KlinkhamerLing2019} is rather complicated.

The goal of the present article is, first, to simplify
the $q$-theory Milne-like model and, second, to study
perturbations near $t=0$.
This last topic is relevant to issues such as stability and
cross-big-bang information transfer.
Precisely these issues have already been discussed
in Ref.~\cite{KlinkhamerWang2019-nonsing-bounce-pert}
for the case of the degenerate-metric nonsingular bounce, but the
case of the Milne-like model is more subtle.

\section{Simplified model for a Milne-like universe}
\label{sec:Simplified-model-for-a-Milne-like-universe}

\subsection{Action and field equations}
\label{subsec:Action and field equations}

The $q$-theory model of a Milne-like universe in Ref.~\cite{KlinkhamerLing2019} uses a postulated dissipative
term in the dynamic equation for the $q$-field.
A relatively simple alternative is to use an action with explicit
derivative terms of the $q$-field~\cite{KV2016-q-ball,%
KV2016-q-DM,KV2016-more-on-q-DM,KlinkhamerMistele2017,Klinkhamer-etal2019}.
Specifically, we take the following
action~\cite{KlinkhamerVolovik2008a,KV2016-q-DM}:
\begin{subequations}\label{eq:action-sigmaLambda-definition-Fq-definition}
 \begin{eqnarray}
S&=&- \int
\,d^4x\, \sqrt{-g}\,\left(\frac{R}{16\pi G_{N}} +\epsilon(q)
\right.
\left.
+\frac{1}{2}\, (q_{0})^{-1}\,
g^{\alpha\beta}\, (\nabla_\alpha\, q)\,(\nabla_\beta\, q)
\right),
\label{eq:action}
\eeqa
\beqa
\label{eq:sigmaLambda-definition}
\epsilon(q) &=&  \sigma(q) + \Lambda\,,
\quad
\frac{d \sigma(q)}{d q}\ne 0\,,
\\[2mm]
\label{eq:Fq-definition}
F_{\alpha\beta\gamma\delta} &\equiv&
\nabla_{[\alpha}A_{\beta\gamma\delta]}\,,
\quad
F_{\alpha\beta\gamma\delta} =
q\,\sqrt{-g} \,\epsilon_{\alpha\beta\gamma\delta}\,,
\end{eqnarray}
\end{subequations}
where $A$ is a three-form gauge field with
corresponding four-form field strength $F \propto q$,
so that the $q$-field has mass dimension $2$.
In \eqref{eq:sigmaLambda-definition},
$\Lambda$ is an initial cosmological constant
and $\sigma(q)$ a generic even function of $q$.
The conventions for the curvature tensors follow those of
Ref.~\cite{Weinberg1972}.

Let us now give a brief review of the resulting field equations.
First, there is the following nonlinear Klein--Gordon
equation~\cite{KV2016-q-DM}:%
\beq\label{eq:nonlin-Klein-Gordon-eq}
(q_{0})^{-1} \, \Box \, q  = \frac{d \rho_{V}(q)}{d q}\,,
\eeq
in terms of the vacuum energy density $\rho_{V}(q)$ defined by
\beq\label{eq:rhoV}
\rho_{V}(q) \equiv \epsilon(q) - \mu_{0}\,q\,,
\eeq
where the constant $\mu_{0}$ traces back to
the solution of the generalized Maxwell equation (the constant $\mu_{0}$ can be interpreted as a chemical
potential~\cite{KlinkhamerVolovik2008a,KlinkhamerVolovik2008b}).
Second, there is the Einstein equation~\cite{KV2016-q-DM},%
\begin{subequations}\label{eq:Einstein-eq}
\beqa
&&
R_{\alpha\beta} - \frac12\, g_{\alpha\beta}\,R = - 8\pi G_{N}\,
T_{\alpha\beta}^{\,\text{(q-vac)}} \,,
\\[2mm]
\label{eq:energy-momentum-tensor-of-q}
&&
T_{\alpha\beta}^{\,\text{(q-vac)}}
=- \left( \rho_{V}(q)
+ \frac12\,(q_{0})^{-1}\,\nabla_\gamma \,q \,\nabla^\gamma q
\right)\,         g_{\alpha\beta}
 +(q_{0})^{-1}\, \nabla_\alpha \, q \, \nabla_\beta \, q \,.
\eeqa
\end{subequations}
The energy-momentum-tensor \eqref{eq:energy-momentum-tensor-of-q}
describes the gravitational effects of the quantum vacuum,
which is here characterized by the $q$-field.

For the simplified theory \eqref{eq:action-sigmaLambda-definition-Fq-definition},
we have that the energy momentum tensor \eqref{eq:energy-momentum-tensor-of-q}
of the \emph{composite} pseudoscalar field $q(x)$ has
the same structure as the one
of a fundamental (pseudo-)scalar field $\phi(x)$.
This implies that previous results for a gravitating
fundamental scalar field $\phi(x)$ carry over,
as long as $\mu_{0}$ has the appropriate dependence
on $\Lambda$ (see, e.g., the discussion of the penultimate paragraph in
Sec.~2 of Ref.~~\cite{Klinkhamer-etal2019}).

\subsection{Ans\"{a}tze}
\label{subsec:Ansaetze}

The metric \textit{Ansatz} is given by
the spatially-hyperbolic ($k=-1$)  Robertson--Walker
metric~\cite{MisnerThorneWheeler2017}  
in terms of comoving spatial
coordinates $\{\chi,\, \vartheta,\, \varphi\}$,
\bsubeqs\label{eq:RW-metric-k-is-minus-1-ranges-coordinates}
\beqa\label{eq:RW-metric-k-is-minus-1}
\hspace*{-2mm}
ds^{2}
&\equiv&
g_{\mu\nu}(x)\, dx^\mu\,dx^\nu
=
- dt^{2}
+ a^{\,2}(t)\,\left[d\chi^2+\sinh^2(\chi) \,
\Big( d\vartheta^2+\sin^2(\vartheta)\, d\varphi^2 \Big)\right]\,,
\\[2mm]
\label{eq:ranges-coordinates}
t &\in& (-\infty,\,\infty)\,,
\quad
\chi \in [0,\,\infty)\,,
\quad
\vartheta \in [0,\,\pi]\,,
\quad
\varphi \in [0,\,2\pi)\,,
\eeqa
\esubeqs
where the infinite range of the cosmic time coordinate $t$
is to be noted.

We will use the Ricci curvature scalar
$R(x) \equiv g^{\nu\sigma}(x)\,g^{\mu\rho}(x)\,R_{\mu\nu\rho\sigma}(x)$
and the Kretschmann curvature scalar
$K(x) \equiv R^{\mu\nu\rho\sigma}(x)\,R_{\mu\nu\rho\sigma}(x)$
as diagnostic tools for the behavior near the big bang with $a(0)=0$.
From the metric \eqref{eq:RW-metric-k-is-minus-1-ranges-coordinates},
these curvature scalars are
\bsubeqs\label{eq:R-K-from-a}
\beqa
\label{eq:R-from-a}
R[a(t)] &=&
6\,\left[\frac{\ddot{a}(t)}{a(t)}
           + \left(\frac{\dot{a}(t)}{a(t)}\right)^2
           -\frac{1}{a^2(t)}\right]\,,
\\[2mm]
\label{eq:K-from-a}
K[a(t)] &=&
12\,\left[\left(\frac{\ddot{a}(t)}{a(t)}\right)^2
           + \left(\left(\frac{\dot{a}(t)}{a(t)}\right)^2
           -\frac{1}{a^2(t)}\right)^2\right] \,,
\eeqa
\esubeqs
where the overdot stands for differentiation with respect to $t$.

The homogenous $q$-field \textit{Ansatz} is given by
\beq\label{eq:homogenous-q}
q=q(t)\,.
\eeq
Following-up on the remark below \eqref{eq:energy-momentum-tensor-of-q},
we observe that the homogeneous  $q$-field has
an energy momentum tensor equal to the one of a perfect fluid
with the following energy density and pressure:
\bsubeqs\label{eq:rho-q-vac-P-q-vac}
\beqa
\label{eq:rho-q-vac}
\rho_\text{q-vac} &=&
\frac{1}{2}\,(q_{0})^{-1}\,\left( \frac{d q}{d t} \right)^2
+ \rho_{V}(q)\,,
\\[2mm]
\label{eq:P-q-vac}
P_\text{q-vac} &=&
\frac{1}{2}\,(q_{0})^{-1}\,\left( \frac{d q}{d t} \right)^2
- \rho_{V}(q)\,,
\eeqa
\esubeqs
for the $q$ function from \eqref{eq:homogenous-q}.
Observe that the Lorentz-invariant behavior
$P_\text{q-vac}=-\rho_\text{q-vac}$ only holds for a
constant $q$-field~\cite{KlinkhamerVolovik2008a}.

\subsection{ODEs}
\label{subsec:ODEs}

We now introduce dimensionless variables ($\hbar=c=8 \pi G_{N} =1$).
Writing $\tau$ for the dimensionless version of the
cosmic time coordinate $t$, we have that
$f(\tau)$ is the dimensionless variable corresponding to $q(t)$,
$r_{V}(\tau)$ the dimensionless variable corresponding to $\rho_{V}(t)$,
and
$p_{V}(\tau)$ the dimensionless variable corresponding to $P_{V}(t)$.
In addition,
$u_{0}$ is the dimensionless constant (chemical potential)
corresponding to $\mu_{0}$ and
$\lambda$ the dimensionless cosmological constant
corresponding to $\Lambda$.

The ordinary differential equations (ODEs) for $a(\tau)$ and $f(\tau)$
are the dimensionless second- and first-order
Friedmann equations and
the dimensionless nonlinear Klein--Gordon equation:
\bsubeqs\label{eq:ODEs-simplified-Milne-like}
\beqa
\label{eq:adotdot-ODE-simplified-Milne-like}
&&\frac{\ddot{a}}{a}
= -\frac{1}{3}\,\Big( \dot{f}^2 - r_{V}  \Big)\,,
\\[2mm]
\label{eq:adot-ODE-simplified-Milne-like}
&&
\left( \frac{\dot{a}}{a}\right)^2 -\frac{1}{a^2}
= \frac{1}{3}\,\left( \frac{1}{2}\,\dot{f}^2 + r_{V} \right)\,,
\\[2mm]
\label{eq:fdotdot-ODE-simplified-Milne-like}
&&\ddot{f} +3\,\left( \frac{\dot{a}}{a}\right) \, \dot{f}
=
- \frac{d}{d f}\, r_{V} \,,
\\[2mm]
\label{eq:rV-def-simplified-Milne-like}
&&r_{V}(f)  =  \epsilon(f) - u_{0}\,f\,,
\eeqa
\esubeqs
where
the overdot now stands for differentiation with respect to $\tau$.
Observe the invariance of these ODEs
 under the following time-reversal transformation:
\bsubeqs
\beqa
\tau &\to& -\tau\,,
\\[2mm]
a(\tau) &\to& -a(-\tau)\,,
\\[2mm]
f(\tau) &\to& f(-\tau)\,,
\eeqa
\esubeqs
with the characteristic odd behavior of the scale factor $a(\tau)$
for the Milne-like universe.
The boundary conditions at $\tau=0$ are taken as follows:
\bsubeqs\label{eq:a-f-fdot-bcs-simplified-Milne-like}
\beqa
\label{eq:a-bcs-simplified-Milne-like}
a(0) &=&     0\,,
\\[2mm]
\label{eq:f-bcs-simplified-Milne-like}
f(0) &=&   f_\text{bb}  \,, \;\;\text{with}\;\; r_{V}(f_\text{bb})>0\,,
\\[2mm]
\label{eq:fdot-bcs-simplified-Milne-like}
\dot{f}(0) &=& 0 \,.
\eeqa
\esubeqs
The model ODEs \eqref{eq:ODEs-simplified-Milne-like}
with boundary conditions \eqref{eq:a-f-fdot-bcs-simplified-Milne-like}
constitute the main result of
Sec.~\ref{sec:Simplified-model-for-a-Milne-like-universe}.

For the numerical calculations later on, we use
the following explicit
energy-density function~\cite{KlinkhamerVolovik2016}:
\beqa
\label{eq:epsilon-Ansatz-simplified-Milne-like}
\epsilon(f) &=& \lambda+ f^2 + 1/f^2\,.
\eeqa
From the equilibrium condition
\beqa
\label{eq:equilibrium condition-Ansatz-simplified-Milne-like}
\epsilon(f) -f\,\frac{d \epsilon(f)}{d f} &=& 0\,,
\eeqa
we get, with the particular function
\eqref{eq:epsilon-Ansatz-simplified-Milne-like},
the following equilibrium value $f_{0}$ (real and positive)
and corresponding chemical potential $u_{0}$\,:
\bsubeqs
\beqa
\label{eq:f0-Ansatz-simplified-Milne-like}
f_{0} &=& \sqrt{\Big(\lambda + \sqrt{12 +\lambda^2}\,\Big)\Big/2}\,,
\\[2mm]
\label{eq:u0-Ansatz-simplified-Milne-like}
u_{0} &=& \left. \frac{d \epsilon(f)}{d f} \right|_{f=f_{0}} =
\frac{2\,{\sqrt{2}}\,\left( 4 + {\lambda}^2 +
      \lambda\,{\sqrt{12 + {\lambda}^2}\,} \right) }{{\left( \lambda +
       {\sqrt{12 + {\lambda}^2}} \right) }^{3/2}}
 \,.
\eeqa
\esubeqs
With the explicit result \eqref{eq:u0-Ansatz-simplified-Milne-like}
for $u_{0}$, we then obtain the dimensionless energy
density \eqref{eq:rV-def-simplified-Milne-like},
which enters the ODEs
\eqref{eq:adotdot-ODE-simplified-Milne-like},
\eqref{eq:adot-ODE-simplified-Milne-like}, and
\eqref{eq:fdotdot-ODE-simplified-Milne-like}.
It can be verified that the following equilibrium properties hold:
\bsubeqs\label{eq:rV-equilibrium-properties}
\beqa
r_{V}(f_{0})&=&0\,,
\\[2mm]
dr_{V}/d f \big|_{f=f_{0}}&=&0\,,
\eeqa
\esubeqs
in addition to having a positive second derivative
corresponding to a positive inverse vacuum compressibility~\cite{KlinkhamerVolovik2008a}.

For completeness, we also give the dimensionless versions of
the quantum-vacuum energy density and pressure from \eqref{eq:rho-q-vac-P-q-vac},
and introduce the corresponding equation-of-state parameter $w$\,:
\bsubeqs\label{eq:r-q-vac-p-q-vac}
\beqa
\label{eq:r-q-vac}
r_\text{q-vac} &=& \frac{1}{2}\,\dot{f}^2 + r_{V}(f)\,,
\\[2mm]
\label{eq:p-q-vac}
p_\text{q-vac} &=& \frac{1}{2}\,\dot{f}^2 - r_{V}(f)\,,
\\[2mm]
\label{eq:w-q-vac}
w_\text{q-vac} &\equiv& p_\text{q-vac}/r_\text{q-vac}\,,
\eeqa
\esubeqs
with $f=f(\tau)$.

\subsection{Analytic results}
\label{subsec:Analytic-results}

For an appropriate fixed value of $f(\tau)$, the two Friedmann ODEs
\eqref{eq:adotdot-ODE-simplified-Milne-like}
and \eqref{eq:adot-ODE-simplified-Milne-like}
with boundary conditions \eqref{eq:a-f-fdot-bcs-simplified-Milne-like}
are solved by the following functions:
\bsubeqs\label{eq:zeroth-order-analytic-solution-simplified-Milne-like}
\beqa
\label{eq:zeroth-order-analytic-a-solution-simplified-Milne-like}
\widetilde{a}(\tau) &=& \frac{1}{\sqrt{r_{V}(f_\text{bb})/3}}\;
\sinh\left(\sqrt{r_{V}(f_\text{bb})/3}\;\tau\right) \,,
\\[2mm]
\label{eq:zeroth-order-analytic-f-solution-simplified-Milne-like}
\widetilde{f}(\tau)  &=& f_\text{bb}\,,
\eeqa
\esubeqs
with $r_{V}(f_\text{bb})>0$. However, the functions \eqref{eq:zeroth-order-analytic-solution-simplified-Milne-like}
do not solve the nonlinear Klein--Gordon equation
\eqref{eq:fdotdot-ODE-simplified-Milne-like},
as its left-hand side vanishes trivially but not
its right-hand side.

The combined solution of the ODEs
\eqref{eq:ODEs-simplified-Milne-like}
with boundary conditions \eqref{eq:a-f-fdot-bcs-simplified-Milne-like}
can be obtained by a perturbative series expansion,
\bsubeqs\label{eq:a-f-series-simplified-Milne-like}
\beqa
\label{eq:a-series-simplified-Milne-like}
a_\text{pert}(\tau) &=&
\tau + c_{3}\,\tau^{3}+ c_{5}\,\tau^{5}+ c_{7}\,\tau^{7}
+ \ldots \,,
\\[2mm]
\label{eq:f-series-simplified-Milne-like}
f_\text{pert}(\tau)  &=& f_\text{bb}\,+
d_{2}\,\tau^{2}+ d_{4}\,\tau^{4}+ d_{6}\,\tau^{6}
+ \ldots \,.
\eeqa
\esubeqs
Defining
\bsubeqs\label{eq:r0-rn-def}
\beqa
\label{eq:r0-def}
r_{0} &\equiv&
r_{V}(f_\text{bb}) > 0\,,
\\[2mm]
\label{eq:rn-def}
r_{n} &\equiv&
\left. \frac{d^{n} r_{V}(f)}{(d f)^{n}} \right|_{f=f_\text{bb}}\,,
\;\;\text{for}\;\; n \geq 1\,,
\eeqa
\esubeqs
the ODEs then give the following results for the coefficients:
\bsubeqs\label{eq:coefficients-cn}
\beqa
\overline{c}_{3} &=&
\frac{r_{0}}{18}\,,
\\[2mm]
\overline{c}_{5} &=&
\frac{8\,r_{0}^2 -
     27\,r_{1}^2}{8640}\,,
\\[2mm]
\overline{c}_{7} &=&
\frac{16\,r_{0}^3 - 54\,r_{0}\,r_{1}^2 +
     405\,r_{1}^2\,r_{2}}{2177280}\,,
\eeqa
\esubeqs
\bsubeqs\label{eq:coefficients-dn}
\beqa
\label{eq:coefficient-d2}
\overline{d}_{2} &=&
-\frac{r_{1}}{8}\,,
\\[2mm]
\overline{d}_{4} &=&
\frac{r_{1}\,
     \left( 2\,r_{0} + 3\,r_{2} \right) }{576}\,,
\\[2mm]
\overline{d}_{6} &=&
-\frac{ r_{1}\,\left( 112\,r_{0}^2 +
         162\,r_{1}^2 + 180\,r_{0}\,r_{2} +
         90\,{r_{2}}^2 + 135\,r_{1}\,r_{3} \right)}{829440}\,.
\eeqa
\esubeqs
For the interpretation of
\eqref{eq:a-series-simplified-Milne-like}
and \eqref{eq:f-series-simplified-Milne-like}
as perturbative series, it is appropriate to consider
$r_{V}$ as the ``small'' perturbation, in line with the structure
of the obtained coefficients
\eqref{eq:coefficients-cn} and \eqref{eq:coefficients-dn}.

With the perturbative solution \eqref{eq:f-series-simplified-Milne-like}
and \eqref{eq:coefficients-dn}, we obtain the following series expansions
of the dimensionless quantum-vacuum energy density $r_\text{q-vac}$
from \eqref{eq:r-q-vac} and pressure $p_\text{q-vac}$ from \eqref{eq:p-q-vac}:
\bsubeqs\label{eq:pert-solution-rV-pV}
\beqa
r_\text{q-vac,\;pert}(\tau)
&=&
r_{0} - \frac{3\,r_{1}^2}{32}\,{\tau}^2 +
  \frac{r_{1}^2\,r_{2}}{128}\,{\tau}^4 + \text{O}(\tau^6)\,,
\\[2mm]
p_\text{q-vac,\;pert}(\tau)
&=&
-r_{0} + \frac{5\,r_{1}^2}{32}\,{\tau}^2 -
  \frac{r_{1}^2\,\left( 8\,r_{0} + 21\,r_{2}
\right)}{1152} \,{\tau}^4  +\text{O}(\tau^6)\,,
\eeqa
\esubeqs
where the cosmological-constant behavior ($r_\text{q-vac}=-p_\text{q-vac}$)
disappears as $\tau$ moves away from $\tau=0$.

With the perturbative solution \eqref{eq:a-series-simplified-Milne-like}
and \eqref{eq:coefficients-cn}, we obtain
the following series expansions of the
dimensionless Ricci curvature scalar $R$
from \eqref{eq:R-from-a}
and the dimensionless Kretschmann curvature scalar $K$
from \eqref{eq:K-from-a}:
\bsubeqs\label{eq:pert-solution-R-and-K}
\beqa\label{eq:pert-solution-R}
R_\text{pert}(\tau) &=&
4\,r_{0} - \frac{9\,r_{1}^2}{16}\,{\tau}^2 +
   \frac{r_{1}^2\,\left( r_{0} + 3\,r_{2} \right)}{48}\, {\tau}^4 + \text{O}(\tau^6)\,,
\\[2mm]
\label{eq:pert-solution-K}
K_\text{pert}(\tau) &=&
\frac{8\,r_{0}^2}{3} -
   \frac{3\,r_{0}\,r_{1}^2}{4}\,{\tau}^2 +
   \frac{r_{1}^2\,\left( 64\,r_{0}^2 +
        135\,r_{1}^2 + 192\,r_{0}\,r_{2} \right) }{2304}\,{\tau}^4 +\text{O}(\tau^6)\,,
\eeqa
\esubeqs
which illustrate the regular behavior at $\tau=0$
of the Milne-like universe.

Our focus in this article is on the behavior near the
big bang at $\tau=0$, but we can also give the asymptotic solution
for $|\tau|\to\infty$. With the near-equilibrium behavior
$r_{V}(f) \propto \big(f-f_{0}\big)^2$,
the ODEs \eqref{eq:ODEs-simplified-Milne-like}
are solved by the following asymptotic solution:
\bsubeqs\label{eq:asymptotic-solution-a-f}
\beqa
a_\text{asymp}(\tau) &\sim&  \tau\,,
\\[2mm]
f_\text{asymp}(\tau) &\sim&  f_{0}\,,
\eeqa
\esubeqs
which corresponds to the Milne
universe~\cite{Milne1932,BirrellDavies1982,Mukhanov2005}
with a nongravitating quantum vacuum in equilibrium.

\begin{figure*}[t]
\vspace*{-0cm}
\begin{center}  
\includegraphics[width=1\textwidth]
{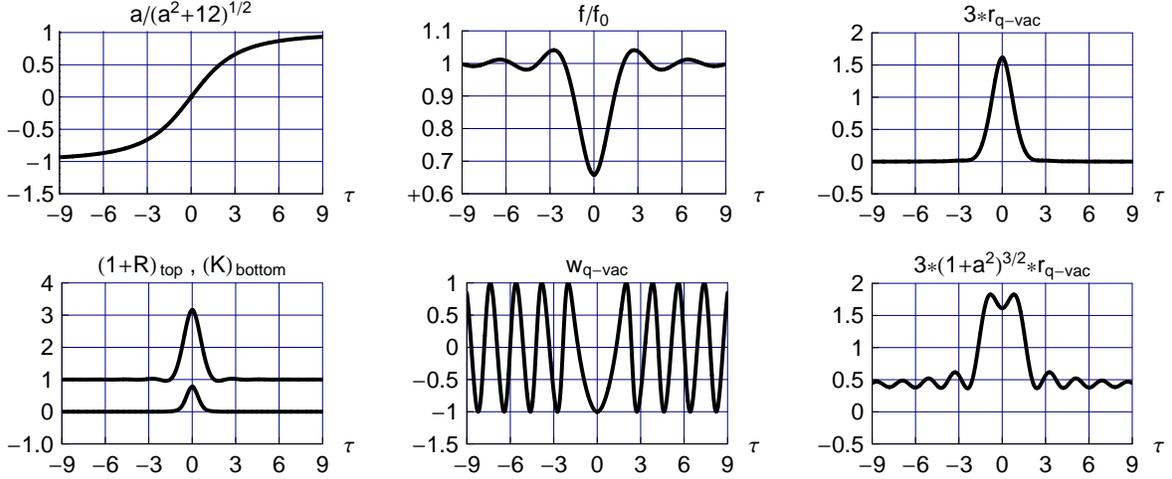}
\end{center}
\vspace*{-4mm}
\caption{Numerical solution of the dimensionless ODEs \eqref{eq:ODEs-simplified-Milne-like} with energy-density
function \eqref{eq:epsilon-Ansatz-simplified-Milne-like} for model parameter $\lambda=1$. The boundary conditions
at $\tau=0$ are $\{a(0),\,f(0),\,\dot{f}(0)\}=\{0,\, 1,\,0\}$,
hence $f_\text{bb}=1$ in the notation of \eqref{eq:a-f-fdot-bcs-simplified-Milne-like}.
The approximate analytic solution for $\tau \in [-\Delta\tau,\,\Delta\tau]$
is given by \eqref{eq:zeroth-order-analytic-solution-simplified-Milne-like}
and the numerical solution is obtained for
$\tau \in [-\tau_\text{max},\,-\Delta\tau]
 \cup   [\Delta\tau,\,\tau_\text{max}]$
with matching boundary conditions at $\tau =\pm \Delta\tau$,
for $\Delta\tau=1/100$  and $\tau_\text{max}=2 \times 10^2$.
Shown, over a reduced time interval,
are the dynamic variables $a(\tau)$ and $f(\tau)$,
together with the corresponding dimensionless quantum-vacuum
energy density $r_\text{q-vac}(\tau)$ from \eqref{eq:r-q-vac},
the equation-of-state parameter $w_\text{q-vac}(\tau)$ from
\eqref{eq:w-q-vac},
the  dimensionless Ricci curvature scalar $R(\tau)$
from \eqref{eq:R-from-a},
and the dimensionless Kretschmann curvature scalar $K(\tau)$
from \eqref{eq:K-from-a}.
The bottom-left panel shows $(1+R)$ as the top curve
and $K$ as the bottom curve.
The bottom-right panel shows $r_\text{q-vac}$ multiplied
by $3\,(1+a^2)^{3/2}$.
The dimensionless $q$-field energy density and pressure
at $\tau=0$ take the following
values: $r_\text{q-vac}(0) \approx 0.537358$
and  $p_\text{q-vac}(0)=-r_\text{q-vac}(0)$.
The dimensionless $q$-field has the Minkowski equilibrium value
$f_{0} \approx 1.51749$ from \eqref{eq:f0-Ansatz-simplified-Milne-like}.
}
\label{fig:simplified-Milne-like-NEAR}
\end{figure*}

For later use, we give an approximate solution
which interpolates between \eqref{eq:a-f-series-simplified-Milne-like}
at lowest order and \eqref{eq:asymptotic-solution-a-f}:
\bsubeqs\label{eq:a-f-approximate-solution}
\beqa
\label{eq:a-approximate-solution}
a_\text{approx}(\tau) &=&\tau\,,
\\[2mm]
\label{eq:f-approximate-solution}
f_\text{approx}(\tau) &=&
\frac{f_\text{bb} +f_{0}\,(\tau/\tau_\text{mid})^2}{1+(\tau/\tau_\text{mid})^2}\,,
\eeqa
\esubeqs
where the optimal value of $\tau_\text{mid}$ needs to be determined
but we will just set $\tau_\text{mid}=1$.

\subsection{Numerical results}
\label{subsec:Numerical-results}

Figure~\ref{fig:simplified-Milne-like-NEAR}
shows numerical
results for the case of $\lambda=1$ and $f_\text{bb}=1$.
As regards the cosmic scale factor $a(\tau)$,
these results are similar to those of
Ref.~\cite{KlinkhamerLing2019}, which employed a source
term to transfer energy from the vacuum sector to the
matter sector. Here, we only have the quantum vacuum, but
temporal fluctuations of the $q$-field
now carry energy which can be red-shifted.
From the results of Fig.~\ref{fig:simplified-Milne-like-NEAR},
we see that the onset of the temporal oscillations occurs at $|\tau| \sim 2$
and that the energy density drops asymptotically as
$r_\text{q-vac} \propto 1/(a^2)^{3/2}$.
Later, we will also consider spatial fluctuations
of the $q$-field, whose energy can, in principle,
be transported away to infinity.

\section{Scalar metric perturbations}
\label{sec:Scalar-metric-perturbations}

\subsection{Background fields}
\label{subsec:Background-fields}

The unperturbed spatially-hyperbolic Robertson--Walker (RW) metric can be rewritten as follows~\cite{MisnerThorneWheeler2017,Mukhanov-etal1992}:
\beq\label{eq:unperturbed-k=-1 conformal metric}
{ds^2}\Big |_{\text{RW}}
=
\Omega^{2} (\eta) \left[ -d {\eta}^2
 + \frac{dx^2+dy^2+dz^2}
 {\big(1-\big[x^2+y^2+z^2\big]\big/4\big)^2}\right] \,,
\eeq
where we have used the conformal time coordinate $\eta$
given by
\beq\label{eq:def.eta}
\Omega(\eta), d\eta \equiv dt \,.
\eeq
Remark that all coordinates $\eta\,,x\,,y$ and $z$ are dimensionless in the metric (\ref{eq:unperturbed-k=-1 conformal metric}),
but that $\Omega (\eta)$ and the corresponding
scale factor $a(t)$ have the dimensions of a length.
For more discussion about this spatially-hyperbolic metric,
see, e.g., the text below Eqs.~(27.25b) and (27.26) in Ref.~\cite{MisnerThorneWheeler2017}.

For the dynamic $q$-field, the homogeneous \textit{Ansatz} is simply
\beq\label{eq:homogeneous-q-field-conformal-coord.}
q(\eta\,,\mathbf{x})= \overline{q}(\eta)\,,
\eeq
where $\overline{q}(\eta)$ is assumed to be nonzero and positive
[the behavior of $\overline{q}(t)$ is shown in the top-mid panel of Fig.~\ref{fig:simplified-Milne-like-NEAR}].
Henceforth, a bar over a quantity denotes its unperturbed (background) value. Logically, $\Omega(\eta)$ or $a(t)$
would also require a bar, but we refrain from doing so,
for the sake of legibility.
	
The energy-momentum tensor from the background $q$-field is
\bsubeqs\label{eq:background-EMT-conformal}
\beqa\label{eq:background-EMT-conformal-00}
\overline{T}^{0}_{\;0} &=&
-\left[\frac{1}{2}\,\frac{(q_{0})^{-1}}{\Omega^{2}}\,
(\overline{q}^{\,\prime})^2+\rho_{V}(\overline{q})\right]\,,
\eeqa\beqa\label{eq:background-EMT-conformal-ij}
\overline{T}^{i}_{\;j} &=&
\left[\frac{1}{2}\,\frac{(q_{0})^{-1}}{\Omega^{2}}\,
(\overline{q}^{\,\prime})^2-\rho_{V}(\overline{q})\right]\,\delta^{i}_{\;j}\,,
\eeqa\beqa\label{eq:background-EMT-conformal-oi}
\overline{T}^{0}_{\;j} &=&0\,,
\eeqa
\esubeqs
where the prime stands for differentiation with respect to $\eta$
and the Latin indices $i$ and $j$ run over $\{1,\,2,\,3\}$.

\subsection{Perturbed fields}
\label{subsec:Perturbed-fields}

In the conformal-Newtonian gauge,
the perturbed spatially-hyperbolic RW metric is given by~\cite{Mukhanov-etal1992}
\beqa\label{eq:perturbed-k=-1_metric}
&&ds^2 \Big | _{\text{RW-perturbed}} ^{\text{(conformal-Newtonian-gauge)}} =
\nonumber\\[1mm]&&
\Omega^{2}(\eta)\, \left[ -(1+2\,\Phi)\,d {\eta}^2
+ \frac{1-2\,\Psi}{\big(1-\big[x^2+y^2+z^2\big]\big/4\big)^2}\;
\big(dx^2+dy^2+dz^2\big) \right] \,,
\eeqa
where the metric perturbations
$\Phi$ and $\Psi$ are functions of all spacetime coordinates
$\{\eta,\, x,\, y,\, z\}$.

For scalar metric perturbations,
the corresponding $q$-field can be decomposed into two parts,
\beq\label{eq:total-q-field-conformal-coord.}
q(\eta\,,\mathbf{x})= \overline{q}(\eta)+\delta q(\eta,\mathbf{x})\,,
\eeq
where $\mathbf{x}$ are the spatial coordinates and $\delta q$ is a small perturbation satisfying $ |\delta q| \ll \overline{q}$,
for nonzero and positive $\overline{q}(\eta)$.
By abuse of notation, we will also write
\beq\label{eq:total-q-field-RW-coord.}
q(t\,,\mathbf{x})= \overline{q}(t)+\delta q(t,\mathbf{x})\,,
\eeq
if, later, we use the original cosmic time coordinate $t$
from \eqref{eq:RW-metric-k-is-minus-1-ranges-coordinates}
and \eqref{eq:def.eta}.

\subsection{Perturbation equations of motion}
\label{subsec:Perturbation equations of motion}

For the perturbed metric (\ref{eq:perturbed-k=-1_metric})
and the perturbed $q$-field \eqref{eq:total-q-field-conformal-coord.},
the first-order perturbation to the energy-momentum tensor
\eqref{eq:energy-momentum-tensor-of-q} is given by
\bsubeqs\label{eq:perturbed-EMT-conformal}
\beqa\label{eq:perturbed-EMT-conformal-00}
\delta T^{0}_{\;0} &=&-\delta \rho_{V} + \frac{ (q_{0})^{-1}}{\Omega^{2}}\,(\overline{q}^{\,\prime})^2\,\Phi
-\frac{(q_{0})^{-1}}{\Omega^{2}}\overline{q}^{\,\prime} \delta q^{\,\prime} \,,
\eeqa\beqa\label{eq:perturbed-EMT-conformal-ij}
\delta T^{i}_{\;j} &=&-\delta^{i}_{\;j}\left[\delta \rho_{V}+\frac{ (q_{0})^{-1}}{\Omega^{2}}\,(\overline{q}^{\,\prime})^2\,\Phi
-\frac{(q_{0})^{-1}}{\Omega^{2}}\overline{q}^{\,\prime} \delta q^{\,\prime}\right]\,,
\eeqa\beqa\label{eq:perturbed-EMT-conformal-oi}
\delta T^{0}_{\;j} &=& -\frac{(q_{0})^{-1}}{\Omega^{2}} \overline{q}^{\,\prime}\,\partial_{j}\, \delta q\,.
\eeqa
\esubeqs
With (\ref{eq:perturbed-EMT-conformal}),
the perturbed Einstein tensor from the metric (\ref{eq:perturbed-k=-1_metric})
gives the following equations of motion for the perturbed
fields (cf. Ref.~\cite{Mukhanov-etal1992}):
\bsubeqs\label{eq:eom-perturbed-q-conformal}
\beqa\label{eq:eom-perturbed-q-conformal-00}
\hspace*{-10mm}
-3\,\mathcal{H}\,\left(\mathcal{H}\,\Phi+\Phi^{\,\prime}\right)
+\nabla^{2}\,\Phi -3\,\Phi
=-4\pi G_{N}\left[ -\Omega ^{2}\,\delta \rho_{V}  +  (q_{0})^{-1}\,(\overline{q}^{\,\prime})^2\,\Phi
-(q_{0})^{-1}\overline{q}^{\,\prime} \,\delta q^{\,\prime}\right],
\eeqa\beqa\label{eq:eom-perturbed-q-conformal-ij}
\hspace*{-10mm}
\left(2\,\mathcal{H}^{\,\prime}+\mathcal{H}^2\right)\,\Phi
+\Phi '' + 3\,\mathcal{H}\,\Phi ' +\Phi
= -4\pi G_{N} \left[ \Omega^{2}\,\delta \rho_{V}
+(q_{0})^{-1}\,(\overline{q}^{\,\prime})^2\,\Phi
-(q_{0})^{-1}\overline{q}^{\,\prime} \,\delta q^{\,\prime}\right],
\eeqa\beqa\label{eq:eom-perturbed-q-conformal-oi}
\hspace*{-10mm}
\partial_{j}\left(\mathcal{H}\,\Phi +\Phi '\right)
=4\pi G_{N}\, (q_{0})^{-1}\, \overline{q}^{\,\prime}\,
\partial_{j}\, \delta q\,,
\eeqa
\esubeqs
with the definition
\beq\label{eq:def.conformal H}
\mathcal{H}\equiv \Omega^{\,\prime}/\Omega
\eeq
and using
\beq\label{eq:Psi-equals-Phi}
\Psi = \Phi\,,
\eeq
as follows from the perturbed off-diagonal spatial Einstein equation.

Remark that
$\nabla ^2$ in \eqref{eq:eom-perturbed-q-conformal-00}
is the Laplace--Beltrami operator on the three-dimensional
constant-curvature hyperbolic space ($\mathbb{H}^3$).
With our spatial $\{x\,,y,\,z\}$ coordinates, this operator
reads explicitly
\beqa
\nabla ^2 \Phi &=& \left(\frac{x^2 +y^2 +z^2}{4}-1 \right)^2\,
\left(\partial^2_{x}\, \Phi +\partial^2_{y}\, \Phi
+\partial^2_{z}\, \Phi\right)
\nonumber\\[1mm]
&&
+\left(\frac{1}{2}-\frac{x^2+y^2+z^2}{8}\right)\,
\left(x\,\partial_x\, \Phi +y\,\partial_y\, \Phi
+ z\,\partial_z\, \Phi\right)\,.
\eeqa
Remark also that, in three-dimensional hyperbolic space, we have the
following eigenvalue equation for $\nabla ^2$\,:
\beq
\nabla ^2\, Q_{k}(\mathbf{x}) = -(k^2 +1) \,Q_{k}(\mathbf{x})\,,
\eeq
where $Q_{k}$ are the eigenfunctions
(with further labels suppressed)
and $-(k^2+1)$ the
eigenvalues~\cite{VilenkinSmorodinsky1964,KodamaSasaki1985,%
CornishSpergel1999}.
In our later discussion, we will take the non-negative real
number $k$ as the physical wavenumber of the $\Phi$ field.

From (\ref{eq:eom-perturbed-q-conformal-00}) and (\ref{eq:eom-perturbed-q-conformal-ij}), we have
the following partial differential equation (PDE):
\beq\label{eq:eom-Phi-conformal-with delta rho}
\Phi'' +6\, \mathcal{H}\, \Phi'
+ 2\,\left( \mathcal{H}'+2\,\mathcal{H}^2+2\right)\, \Phi
- \nabla^{2}\,\Phi
=-8 \pi G_{N} \, \Omega ^{2} \,\delta \rho_{V}\,.
\eeq
Notice that, with the conformal time coordinate $\eta$,
the background nonlinear Klein--Gordon equation
\eqref{eq:nonlin-Klein-Gordon-eq} is given by
\beq\label{eq:K_G eq.-conformal}
\overline{q}^{\,\prime\prime}
+2\,\mathcal{H}\,\overline{q}^{\,\prime}
+ q_{0} \,\Omega^{2}\, \frac{d \rho_{V}(q)}{d q}=0\,,
\eeq
so that we have
\beq\label{eq:expression-delta-rho-with-eta}
\Omega^{2}\, \delta \rho_{V} =
\Omega^{2} \,\frac{d \rho_{V}(q)}{d q}\, \delta q=
- \frac{\overline{q}^{\,\prime\prime}
+2\,\mathcal{H}\,\overline{q}^{\,\prime}}{\overline{q}^{\,\prime}}
\;\frac{\mathcal{H}\, \Phi +\Phi'}{4\pi G_{N}} \,,
\eeq
where we have used (\ref{eq:eom-perturbed-q-conformal-oi}) to express $\delta q$ in terms of $\Phi$ and $\Phi '$\,,
\beq\label{eq:eom-delta q in terms of Phi}
\delta q =
(4\pi G_{N})^{-1} q_{0}\,
\frac{\mathcal{H}\,\Phi +\Phi '}{\overline{q}^{\,\prime}}\,.
\eeq
With (\ref{eq:expression-delta-rho-with-eta}),
the PDE
(\ref{eq:eom-Phi-conformal-with delta rho}) can be written as
\beq\label{eq:eom-Phi-conformal}
\Phi ''
+\,2\left( \mathcal{H} -\frac{\overline{q}^{\,\prime\prime}}{\overline{q}^{\,\prime}}\right) \Phi '+ 2\,\left( \mathcal{H}'-
\frac{\mathcal{H}\, \overline{q}^{\,\prime\prime}}{\overline{q}^{\,\prime}}+2\right)\, \Phi - \nabla^{2}\,\Phi  =0\,,
\eeq
which is the final equation of motion for the metric perturbation
$\Phi(\eta,\,\mathbf{x})$.

Returning to the original cosmic
time coordinate $t$ from \eqref{eq:RW-metric-k-is-minus-1},
we rewrite (\ref{eq:eom-Phi-conformal})
as follows for $\Phi(t,\,\mathbf{x})$:
\beq\label{eq:eom-Phi-with-t}
\ddot{\Phi}+
\left(\frac{\dot{a}}{a}-\frac{2\,\ddot{\overline{q}}}{\dot{\overline{q}}} \right)\dot{\Phi}
+2\left(\frac{\ddot{a}}{a}
-\frac{\dot{a}^2}{a^2}-
\frac{\dot{a}}{a}\,
\frac{\ddot{\overline{q}}}{\dot{\overline{q}}}
+\frac{2}{a^2} \right)\Phi -\frac{\nabla^{2}\,\Phi}{a^2}=0\,,
\eeq
where the overdot stands for the partial derivative with respect to $t$.
Writing the solution of \eqref{eq:eom-Phi-with-t} as
$\widehat{\Phi}(t,\,\mathbf{x})$
and using \eqref{eq:eom-delta q in terms of Phi},
the $q$-field perturbation solution $\widehat{\delta q}(t,\,\mathbf{x})$
is given by
\beq\label{eq:eom-delta q in terms of Phi-with-t}
\widehat{\delta q}/q_{0} =
2\,\frac{(E_\text{planck})^2}{q_{0}}\;
\frac{H\,\widehat{\Phi} +\dot{\widehat{\Phi}}}
{\dot{\overline{q}}/q_{0}}\,,
\eeq
in terms of the reduced Planck energy
\beq\label{eq:E-planck}
E_\text{planck} \equiv \sqrt{1/(8\pi\, G_N)}
\approx 2.44 \times 10^{18}\,\text{GeV}\,.
\eeq
Equations \eqref{eq:eom-Phi-with-t}
and \eqref{eq:eom-delta q in terms of Phi-with-t}
constitute the main general result of
Sec.~\ref{sec:Scalar-metric-perturbations}.

\subsection{Perturbation solutions}
\label{subsec:Perturbation-solutions}

From now on, we use the dimensionless variables introduced in the
first paragraph of Sec.~\ref{subsec:ODEs} and
consider only a spherical-wave scalar
metric perturbation $\Phi$
and a spherical-wave perturbation  $\delta f$ of the dimensionless
$q$-field,
\bsubeqs\label{eq:spherical-wave-Phi-f}
\beqa
\Phi(\tau,\,\mathbf{x})
&=&
\int_{0}^{\infty} d k\, \sum_{l,m}\;
\Phi_{klm}(\tau)\; Q_{klm}(\chi,\, \theta,\, \phi)\,,
\\[2mm]
\delta f(\tau,\,\mathbf{x})
&=&
\int_{0}^{\infty} d k\, \sum_{l,m}\;
\delta f_{klm}(\tau)\; Q_{klm}(\chi,\, \theta,\, \phi)\,,
\eeqa
\esubeqs
where $\{\chi,\, \theta,\, \phi\}$ are the dimensionless
coordinates from \eqref{eq:RW-metric-k-is-minus-1-ranges-coordinates}
and $Q_{klm}(\chi,\, \theta,\, \phi)$ the eigenmodes of the
hyperbolic Laplace operator~\cite{CornishSpergel1999}.
For the labels of the eigenmodes,
we prefer to use the wavenumber $k$ with range
\beq
k\in [0,\,\infty)\,,
\eeq
as we already have used the
symbol $q$ for the physical $q$-field of our model.
Moreover, we take, for definiteness,
the following $q$-field equilibrium value:
\beq\label{eq:q0-is-Eplanck2}
q_{0} = \left(E_\text{planck}\right)^2 \,.
\eeq
From \eqref{eq:eom-Phi-with-t}
and \eqref{eq:eom-delta q in terms of Phi-with-t},
the corresponding equations are
\bsubeqs\label{eq:eom-Phikamplitude-with-t-deltafkamplitude-solution}
\beqa\label{eq:eom-Phikamplitude-with-t}
\hspace*{-10mm}
&&\ddot{\Phi}_{klm}+
\left(\frac{\dot{a}}{a}
-\frac{2\,\ddot{\overline{f}}}{\dot{\overline{f}}}\right)
\dot{\Phi}_{klm}
+2\left(\frac{\ddot{a}}{a}
- \frac{\dot{a}^2}{a^2}-
\frac{\dot{a}}{a}\,\frac{\ddot{\overline{f}}}{\dot{\overline{f}}}
+\frac{2}{a^2} \right)\Phi_{klm} + \left(1+k^2\right)\,\frac{\Phi_{klm}}{a^2}=0\,,
\\[2mm]
\label{eq:deltafkamplitude-solution}
\hspace*{-10mm}
&&\widehat{\delta f}_{klm}/f_{0} =
2\;
\frac{h\,\widehat{\Phi}_{klm} +\dot{\widehat{\Phi}}_{klm}}
{\dot{\overline{f}}/f_{0}}\,,
\eeqa
\esubeqs
with $-(1+k^2)$ the eigenvalue of the hyperbolic Laplace
operator~\cite{VilenkinSmorodinsky1964,KodamaSasaki1985,CornishSpergel1999},
the dimensionless Hubble parameter $h \equiv \dot{a}/a$,
and the overdot now standing for the partial derivative
with respect to $\tau$.

The linear ODE~\eqref{eq:eom-Phikamplitude-with-t}
is singular, because the
background cosmic scale factor $a(\tau)$ vanishes at $\tau=0$,
according to the boundary condition \eqref{eq:a-bcs-simplified-Milne-like}.
In the present article, we present some exploratory
results for this singular ODE.

\subsubsection{Absence of a regular solution at $\tau=0$}
\label{subsubsec:Absence-regular-solution}

Consider the following truncated series for the
background functions:
\bsubeqs\label{eq:a-f-truncated-series-simplified-Milne-like}
\beqa
\label{eq:a-truncated-series-simplified-Milne-like}
a^\text{(trunc-series)}(\tau) &=&
\tau + \sum_{n=1}^{N}\,c_{2n+1}\,\tau^{2n+1}\,,
\\[2mm]
\label{eq:f-truncated-series-simplified-Milne-like}
\overline{f}^\text{(trunc-series)}(\tau)  &=& f_\text{bb}\,
+ \sum_{n=1}^{N}\,d_{2n}\,\tau^{2n} \,,
\eeqa
\esubeqs
where analytic results for the coefficients $c_{2n+1}$ and $d_{2n}$
have been presented in Sec.~\ref{subsec:Analytic-results}.
Also consider a  truncated series for the scalar-perturbation
amplitude,
\beq\label{eq:Phikamplitude-truncated-series}
\Phi_{klm}^\text{(trunc-series)}(\tau)
=\sum_{n=1}^{N'}\,p_{klm,\,2n}\,\tau^{2n}\,.
\eeq

Inserting \eqref{eq:a-f-truncated-series-simplified-Milne-like}
and \eqref{eq:Phikamplitude-truncated-series}
into the ODE \eqref{eq:eom-Phikamplitude-with-t},
we obtain that all coefficients $p_{klm,\,2n}$
vanish if $N$ and $N'$ are taken larger and larger,
\beq\label{eq:coeff-p-2n-vanishing}
p_{klm,\,2n}=0, \;\;\text{for}\;\; n=1,\,2,\,3,\, \ldots \,.
\eeq
The result \eqref{eq:coeff-p-2n-vanishing}
suggests that the scalar-perturbation amplitude $\Phi_{klm}(\tau)$
has an essential singularity at $\tau=0$.
From \eqref{eq:deltafkamplitude-solution}, the same is expected
to hold for the $q$-field perturbation amplitude
$\delta f_{klm}(\tau)$.

\subsubsection{Essential singularity at $\tau=0$}
\label{subsubsec:Essential-singularity}

For certain simplified background functions, we
are able to obtain the exact solutions
$\Phi_{klm}(\tau)$ and $\delta f_{klm}(\tau)$.
These exact solutions display an essential singularity at $\tau=0$,
consistent with the results of
Sec.~\ref{subsubsec:Absence-regular-solution}.

Consider the following simplified background functions:
\bsubeqs\label{eq:a-f-simplified-Milne-like}
\beqa
\label{eq:a-simplified-Milne-like}
a^\text{(simplified-background)}(\tau) &=&\tau \,,
\\[2mm]
\label{eq:f-simplified-Milne-like}
\overline{f}^\text{(simplified-background)}(\tau)  &=& f_\text{bb}\,+ d_{2}\,\tau^{2} \,,
\eeqa
\esubeqs
with a constant $d_{2}$ which may or may not be related
to \eqref{eq:coefficient-d2}  in Sec.~\ref{subsec:Analytic-results}.
Inserting \eqref{eq:a-f-simplified-Milne-like}
into the ODE \eqref{eq:eom-Phikamplitude-with-t},
we obtain the following analytic solution
\bsubeqs\label{eq:a-f-simplified-Milne-like-Phi-solution-deltaf-solution}
\beq\label{eq:a-f-simplified-Milne-like-Phi-solution}
\Phi_{klm}^\text{(simplified-background)}(\tau) =
C_{klm}\,{\sqrt{{\tau}^2}}\,
  \cos \left[\alpha_{klm} + \frac{1}{2}\,k\,\log \left({\tau}^2\right)\right]\,,
\eeq
where $C_{klm}$ and $\alpha_{klm}$ are real constants.
With \eqref{eq:deltafkamplitude-solution}, there is the
corresponding scalar perturbation,
\beqa\label{eq:a-f-simplified-Milne-like-deltaf-solution}
\hspace*{-10mm}
&&\delta f_{klm}^\text{(simplified-background)}(\tau)/f_{0} =
C_{klm}\;
\frac{ 2\,\cos \left[\alpha_{klm} + \frac{1}{2}\,k\,\log \left({\tau}^2\right)\right] -
      k\,\sin \left[\alpha_{klm} + \frac{1}{2}\,k\,\log \left({\tau}^2\right)\right]}   {d_{2}\,{\sqrt{{\tau}^2}}}\,.
\nonumber\\\hspace*{-10mm}&&
\eeqa
\esubeqs
The analytic solutions
\eqref{eq:a-f-simplified-Milne-like-Phi-solution}
and \eqref{eq:a-f-simplified-Milne-like-deltaf-solution}
are the most important concrete results of Sec.~\ref{sec:Scalar-metric-perturbations}.

\begin{figure*}[t]
\vspace*{-0cm}
\begin{center} 
\includegraphics[width=1\textwidth]
{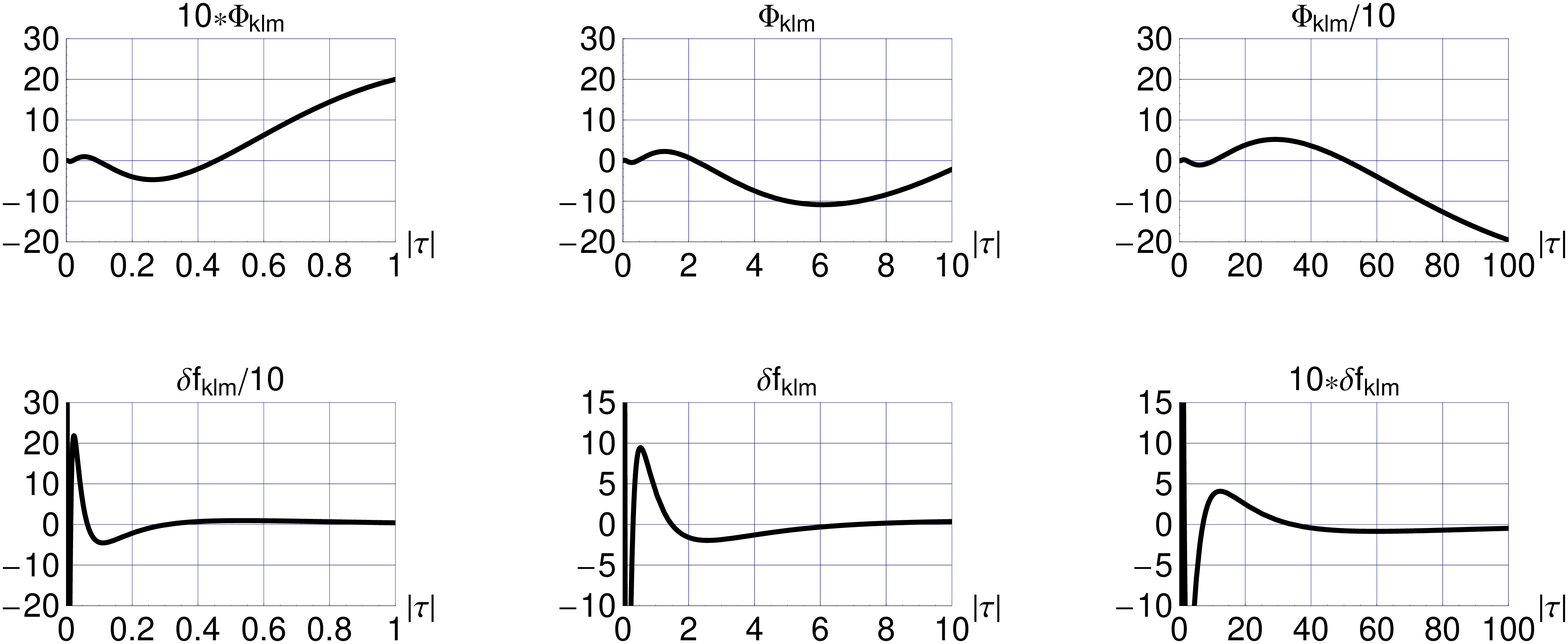}
\end{center}
\vspace*{-4mm}
\caption{Analytic solutions \eqref{eq:a-f-simplified-Milne-like-Phi-solution-deltaf-solution}
for the perturbations $\Phi_{klm}(\tau)$ and $\delta f_{klm}(\tau)$,
using the simplified background functions $a(\tau)$ and $f(\tau)$
from \eqref{eq:a-f-simplified-Milne-like}.
The parameters are
$\{k,\, C_{klm},\,\alpha_{klm},\, f_\text{bb},\,d_2\}
=\{2,\, 2,\, 0,\, 1,\, 1\}$.
}
\label{fig:perturbations-analytic-solution}
\end{figure*}

Both functions \eqref{eq:a-f-simplified-Milne-like-Phi-solution}
and \eqref{eq:a-f-simplified-Milne-like-deltaf-solution}
display the characteristic behavior of an essential singularity:
rapidly increasing oscillations as $\tau^2$ approaches the value $0$.
The  scalar perturbation
\eqref{eq:a-f-simplified-Milne-like-deltaf-solution} has,
moreover, an increasing amplitude as $\tau^2$ approaches $0$
(see Fig.~\ref{fig:perturbations-analytic-solution}).

\subsubsection{Numerical results}
\label{subsubsec:Numerical-results}

For a preliminary numerical calculation, we
use the following simplified background functions
which have the correct global behavior:
\bsubeqs\label{eq:a-f-global-simplified-Milne-like}
\beqa
\label{eq:a-global-simplified-Milne-like}
a^\text{(global-simplified-background)}(\tau) &=&\tau \,,
\\[2mm]
\label{eq:f-global-simplified-Milne-like}
\overline{f}^\text{(global-simplified-background)}(\tau)
&=& \frac{f_\text{bb}\,+ f_{0}\,\tau^{2}}{1+ \tau^{2}} \,.
\eeqa
\esubeqs
We then get the numerical $\Phi_{klm}(\tau)$ solution from
\eqref{eq:eom-Phikamplitude-with-t} with
boundary conditions at $\tau=1$.
For the sake of comparison, these boundary conditions at $\tau=1$
are taken to match the previous analytic function
\eqref{eq:a-f-simplified-Milne-like-Phi-solution}
with $C_{klm}=2$ and $\alpha_{klm}=0$.
Inserting that numerical  $\Phi_{klm}(\tau)$ solution in
\eqref{eq:deltafkamplitude-solution} gives
the dimensionless $q$-field perturbation $\delta f_{klm}(\tau)$.

\begin{figure*}[t]
\vspace*{-0cm}
\begin{center}
\includegraphics[width=1\textwidth]
{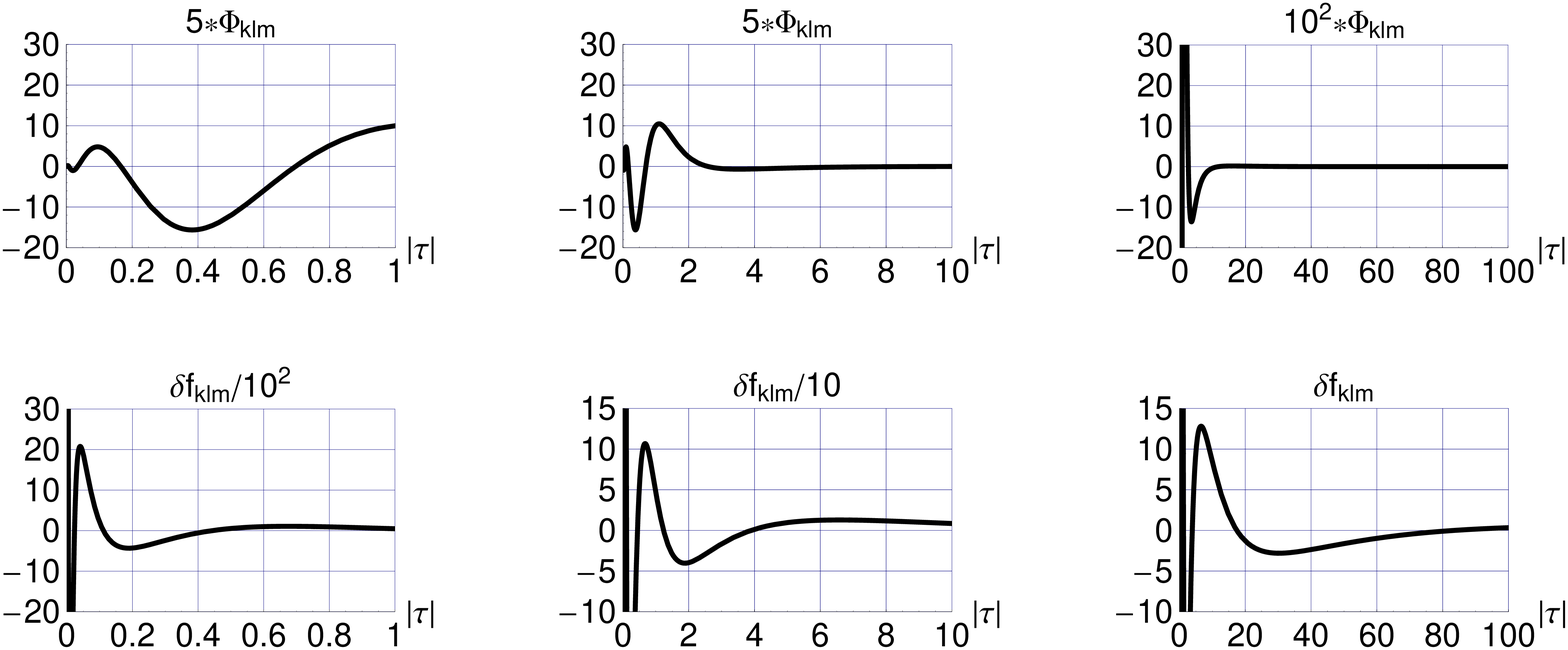}
\end{center}
\vspace*{-4mm}
\caption{Numerical solution $\Phi_{klm}(\tau)$ from the
ODE \eqref{eq:eom-Phikamplitude-with-t}
with boundary conditions
$\Phi_{klm}(1)=2$ and $\dot{\Phi}_{klm}(1)=2$,
for background functions $a(\tau)$ and $f(\tau)$ from \eqref{eq:a-f-global-simplified-Milne-like}.
The dimensionless $q$-field perturbation $\delta f_{klm}(\tau)$ follows
from the numerical $\Phi_{klm}(\tau)$ solution and
\eqref{eq:deltafkamplitude-solution}. The parameters are
$\{k,\, f_\text{bb},\,f_0\} =\{2,\, 1,\, 1.51749\}$.
}
\label{fig:perturbations-numerical-solution-global-simplified}
\vspace*{15mm}
\vspace*{-0cm}
\begin{center}
\includegraphics[width=1\textwidth]
{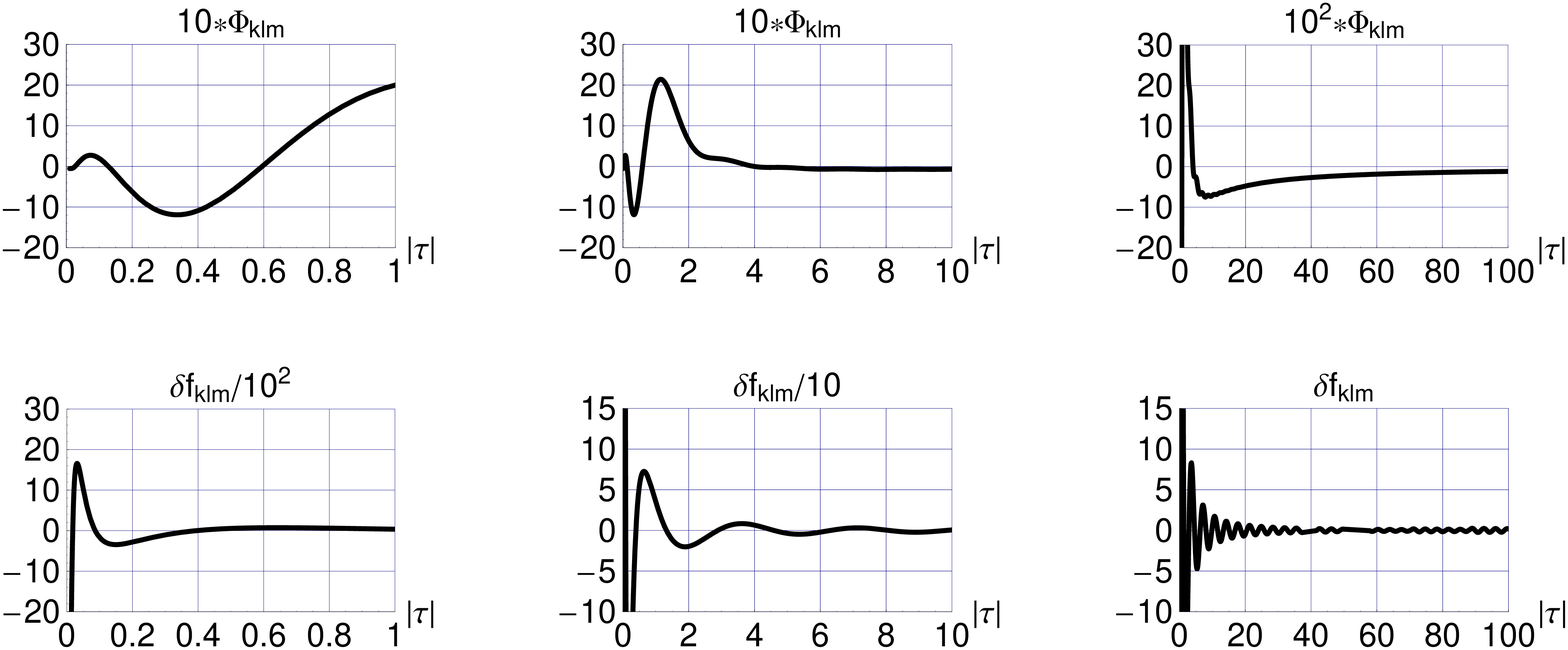}
\end{center}
\vspace*{-4mm}
\caption{Numerical solution $\Phi_{klm}(\tau)$ from the
ODE \eqref{eq:eom-Phikamplitude-with-t}
with boundary conditions
$\Phi_{klm}(1)=2$ and $\dot{\Phi}_{klm}(1)=2$,
for numerical background functions $a(\tau)$ and $f(\tau)$ from
Fig.~\ref{fig:simplified-Milne-like-NEAR}.
The dimensionless $q$-field perturbation $\delta f_{klm}(\tau)$ follows
from the numerical $\Phi_{klm}(\tau)$ solution and
\eqref{eq:deltafkamplitude-solution}. The parameters are
$\{\lambda,\, k,\, f_\text{bb},\,f_0\} =\{1,\,2,\, 1,\, 1.51749\}$.
}
\label{fig:perturbations-numerical-solution}
\end{figure*}

These preliminary numerical results are shown in
Fig.~\ref{fig:perturbations-numerical-solution-global-simplified}.
The qualitative behavior for $\tau \lesssim 1$ of
Fig.~\ref{fig:perturbations-numerical-solution-global-simplified}
is similar to that of Fig.~\ref{fig:perturbations-analytic-solution}.
New in Fig.~\ref{fig:perturbations-numerical-solution-global-simplified}
is that the $\Phi_{klm}$ oscillations essentially disappear for
larger $\tau$ values, most likely, because the $\overline{f}(\tau)$
function from \eqref{eq:f-global-simplified-Milne-like} approaches its asymptotic equilibrium value $f_0$,
instead of growing indefinitely
as in \eqref{eq:f-simplified-Milne-like}.

For the final numerical calculation, we use the numerical
background functions $a(\tau)$ and $\overline{f}(\tau)$
from Fig.~\ref{fig:simplified-Milne-like-NEAR}
to get the perturbations
$\Phi_{klm}(\tau)$ and $\delta f_{klm}(\tau)$
from \eqref{eq:eom-Phikamplitude-with-t-deltafkamplitude-solution}.
These numerical results are shown in
Fig.~\ref{fig:perturbations-numerical-solution}.

The qualitative behavior for $\tau \lesssim 10$ of
Fig.~\ref{fig:perturbations-numerical-solution} is similar to that of Fig.~\ref{fig:perturbations-numerical-solution-global-simplified}.
New in Fig.~\ref{fig:perturbations-numerical-solution}
are the small $\delta f_{klm}(\tau)$ oscillations
starting at $\tau \sim 2$, which are, most likely, due to the
oscillations of the background field $\overline{f}(\tau)$
as it approaches the equilibrium value $f_0$
(see Fig.~\ref{fig:simplified-Milne-like-NEAR}).

\section{Discussion}
\label{sec:Discussion}

In this article, we have, first,
presented a simplified model for a time-symmetric Milne-like universe.
In this simplified dynamic-vacuum-energy model,
like in the previous model of Ref.~\cite{KlinkhamerLing2019},
the big bang singularity is just a coordinate
singularity~\cite{Ling2018} with finite curvature and energy density.
We have, then,  calculated the dynamics of scalar metric perturbations.

The perturbation results of
Fig.~\ref{fig:perturbations-numerical-solution}
can be described as follows:
starting with small perturbations
$\Phi_{klm}$ and $\delta f_{klm}$ at $\tau=-100$
and evolving them towards $\tau=0^{-}$, it is found
that the $\delta f_{klm}$ oscillations get a larger
and larger amplitude, so that perturbation theory breaks down
(see App.~\ref{app:Perturbed Ricci scalar} for related results
on the perturbed Ricci curvature scalar).
In short, the initial perturbations from $\tau=-100$
evolve and upset the unperturbed big bang solution $\tau=0^{-}$.
Incidentally, with such a major disruption of the original
big bang behavior, the discussion of
possible cross-big-bang information transfer becomes moot.

At this moment, it may be worthwhile to compare
our perturbation results of the  Milne-like universe
with those of the nonsingular
bounce~\cite{KlinkhamerWang2019-nonsing-bounce-pert}.
Specifically, the results for the scalar metric
perturbations and the corresponding adiabatic perturbations of the
nonrelativistic-matter energy density
were found to be given by
\bsubeqs\label{eq:perturbation-results-nonsingular-bounce}
\beqa
\label{eq:perturbation-results-nonsingular-bounce-Phi-k}
\Phi_\mathbf{k}(t)\,\Big|^\text{(nonsingular-bounce)}
&=&
\widehat{C}_{\mathbf{k},\,1}
+
\frac{b^{5/3}\,\widehat{C}_{\mathbf{k},\,2}}
     {\big(b^{2}+t^{2}\big)^{5/6}}\,,
\\[2mm]
\label{eq:perturbation-results-nonsingular-bounce-deltarho-k}
\frac{\delta \rho_\mathbf{k}(t)}{\bar{\rho}(t)}\,
\Big|^\text{(nonsingular-bounce)}
&=&
-\left[ 2+\frac{3}{2}\, |\mathbf{k}|^{2}\,
\big(b^{2}+t_{0}^{2}\big)^{2/3}\,\big(b^{2}+t^{2}\big)^{1/3}  \right]\widehat{C}_{\mathbf{k},\,1}
\nonumber\\[1mm]
&&
+\left[ 3-\frac{3}{2}\,|\mathbf{k}|^{2}\,
\big(b^{2}+t_{0}^{2}\big)^{2/3}\,\big(b^{2}+t^{2}\big)^{1/3} \right] \frac{b^{5/3}\,\widehat{C}_{\mathbf{k},\,2}}{\big(b^{2}+t^{2}\big)^{5/6}}\,,
\eeqa
\esubeqs
where $b>0$ is the spacetime-defect length scale entering the
metric \textit{Ansatz} and $t_0$ an arbitrary reference time used for
the background cosmic scale factor;
see Ref.~\cite{KlinkhamerWang2019-nonsing-bounce-pert}
for further details.
The behavior of the perturbations
\eqref{eq:perturbation-results-nonsingular-bounce}
at the moment of the bounce, $t=0$,
is perfectly regular due to the presence of the $b^2$ terms in the
various denominators.
This regular behavior of the nonsingular-bounce
perturbations
\eqref{eq:perturbation-results-nonsingular-bounce} contrasts with
the essential-singularity
behavior of the Milne-like-universe perturbations as shown in
\eqref{eq:a-f-simplified-Milne-like-Phi-solution-deltaf-solution}.

The heuristic explanation for the different behavior
discussed in the
previous paragraph may be twofold. First, the nonsingular bounce
has a nonvanishing cosmic scale factor $a(t)$ at the bounce,
whereas the Milne-like universe has a
vanishing cosmic scale factor $a(t)$ at the big bang,
which requires delicate cancellations in order to
give a mere coordinate singularity and these cancellations
may be destroyed by perturbations of the metric.
Second, the nonsingular bounce has a ``hard-wired'' parameter
$b$ to regularize the potential big bang singularity,
whereas the Milne-like universe has a positive
vacuum energy density, which traces back
to a dynamic vacuum energy density $\rho_{V}(q)$.

This last problematic point of the Milne-like model
would disappear if the variable quantity
$\rho_{V}(q)$ were replaced by a positive
cosmological constant $\Lambda$, but then the question
resurfaces as how to remove the nonvanishing
vacuum energy density as the universe evolves.
More attractive would be to keep
the dynamic vacuum energy density $\rho_{V}(q)$ and
to find a mechanism that \emph{freezes} $q(x)$ at $t=0$,
which then forces having a vanishing perturbation
$\delta q(x)=0$ at $t=0$
[this would effectively set $C_{klm}=0$ in
\eqref{eq:a-f-simplified-Milne-like-Phi-solution-deltaf-solution}].
Such a freezing of the $q$-field
has been discussed in a different context (cf. Sec.~IV of Ref.~\cite{KlinkhamerVolovik2009}) and may indeed be the preferred way
to rescue the big bang coordinate singularity of the Milne-like universe.

\vspace*{0mm}
\begin{acknowledgments}
\vspace*{-2mm}
The work of Z.L.W. is supported by the China Scholarship Council.
\end{acknowledgments}

\begin{appendix}
\section{Perturbed Ricci scalar}
\label{app:Perturbed Ricci scalar}

The Ricci curvature scalar from the perturbed spatially-hyperbolic Robertson--Walker metric \eqref{eq:perturbed-k=-1_metric}
is given by:
\beq
R=\frac{1}{\Omega ^{2} (\eta) }
\left[ 6\, \left(\frac{\Omega''}{\Omega}-1 \right) -54\, \Phi'' -24\, \frac{\Omega '}{\Omega}\,\Phi ' -12\,\frac{\Omega ''}{\Omega}\Phi -12\,\Phi -46\,\nabla ^2\, \Phi \right]
+\text{O}\left(\Phi ^2\right),
\eeq
where the prime stands for differentiation with respect to $\eta$
and where \eqref{eq:Psi-equals-Phi} has been used.
Changing to the original cosmic time coordinate $t$ from
\eqref{eq:def.eta} and setting $\Omega(\eta)=a(t)$,
we have
\beq
R=6\,\left( \frac{\ddot{a}}{a}
+\frac{\dot{a} ^2}{a^2}-\frac{1}{a^2}\right) -54\,\ddot{\Phi}-78\,\frac{\dot{a}}{a}\dot{\Phi} -12\, \left(\frac{\dot{a}^2}{a^2}+\frac{\ddot{a}}{a} \right)\Phi -12\,\frac{\Phi}{a^2} -46\, \frac{\nabla ^2 \Phi}{a^2}
+\text{O}\left(\Phi ^2\right),
\eeq
where the overdot stands for differentiation with respect to $t$.

With the background function \eqref{eq:a-simplified-Milne-like}
and the analytic $\Phi$ solution
\eqref{eq:a-f-simplified-Milne-like-Phi-solution},
we then get the following first-order perturbation
of the dimensionless Ricci curvature scalar
($\tau$ is the dimensionless version of the time coordinate $t$):
\beqa\label{eq:deltaR-result}
\delta R_{klm}(\tau)&=&
\frac{C_{klm}}{\sqrt{\tau ^2}}\Bigg( \left(100\, k^2-44\right)\,
\cos \left[\alpha_{klm} + \frac{1}{2}\,k \,\log\left(\tau ^2\right)\right]
\nonumber\\[1mm]
&&
+132 \,k \,\sin\left[\alpha_{klm} + \frac{1}{2}\,k\, \log\left(\tau ^2\right)\right]  \Bigg)\,,
\eeqa
which displays
the same type of behavior as found for $\delta f_{klm}(\tau)$ in
Sec.~\ref{subsec:Perturbation-solutions}.
In fact, the perturbation \eqref{eq:deltaR-result}
ruins the regular behavior at $\tau=0$ of the
background Ricci scalar \eqref{eq:pert-solution-R}.

\end{appendix}


\end{document}